\documentclass[aps,preprint,amssymb,12pt,floatfix]{revtex4}
\usepackage{subfigure}
\usepackage{graphicx}
\usepackage{rotating}

\begin{document}

\title{Effect of finite size on cooperativity and rates of protein folding}
\author{Maksim Kouza$^1$, Mai Suan Li$^1$, Edward P. O'Brien Jr.$^2$,
Chin-Kun Hu$^{3,4}$ and D. Thirumalai$^{2,5}$}

\address{$^1$Institute of Physics, Polish Academy of Sciences,
Al. Lotnikow 32/46, 02-668 Warsaw, Poland\\
$^2$Biophysics Program, Institute for Physical
Science and Technology, University of Maryland, College Park, MD 20742\\
$^3$Institute of Physics, Academia Sinica, Nankang,
Taipei 11529, Taiwan\\
$^4$National Center for Theoretical Sciences at
Taipei, Physics Division, National Taiwan University, Taipei
10617, Taiwan\\
$^5$Department of Chemistry and Biochemistry, University of Maryland, 
College Park, MD 20742}

\date{\small \today}

\baselineskip = 22pt

\begin{abstract}

We analyze the dependence of 
cooperativity of the thermal denaturation transition
and folding rates of globular proteins on the number of amino acid
residues, $N$, using lattice models with side chains,
off-lattice Go models and the 
available experimental data. A dimensionless
measure of cooperativity, $\Omega _c$ ( $0 < \Omega _c < \infty$),
scales as $\Omega_c \sim N^{\zeta}$. The results of simulations and
the analysis of experimental data further confirm the earlier prediction
that $\zeta$ is universal with $\zeta = 1 +\gamma$, where
exponent $\gamma$ characterizes the susceptibility of a self-avoiding
walk. This finding suggests that the structural
characteristics in
the denaturated state are manifested in the folding cooperativity
at the transition temperature.
The folding
rates $k_F$ for the Go models and a dataset of 69 proteins can be fit 
using $k_F =  k_F^0 \exp(-cN^\beta)$.
Both $\beta = 1/2$ and $2/3$ provide a good fit of the data.
We find that $k_F =  k_F^0 \exp(-cN^{\frac{1}{2}})$, with
the average (over the dataset of proteins)
$k_F^0 \approx (0.2\mu s)^{-1}$ and $c \approx 1.1$,
can be used to estimate folding rates to within an order
of magnitude in most cases. The minimal models give identical $N$ 
dependence with $c \approx 1$. The prefactor for off-lattice Go models
is nearly four
orders of magnitude larger than the experimental value.
\end{abstract}

%\pacs{PACS numbers: 87.10.+e,87.15.Cc,36.20.-r}

%]
\maketitle

\section{Introduction}

Single domain globular proteins are mesoscopic systems that self-assemble,
under folding conditions, to a compact state with definite topology.  Given
that the folded states of proteins are only on the order of tens of
Angstroms
(the radius of gyration $R_g \approx 3 N^{\frac {1}{3}}$ \AA $~$
\cite{DimaThirJPCB04} where $N$ is
the number of amino acids) it is surprising that they undergo highly
cooperative transitions from an ensemble of unfolded states to the native state 
\cite{Poland,Privalov79}.
Similarly, there is a wide spread in the folding
times as well \cite{Galzitskaya03,Ivankov03,Ivankov04}.
The rates of folding vary by nearly nine orders of
magnitude.  Sometime ago it was shown theoretically that the folding time
,$\tau_F$, should depend on $N$
\cite{Thirumalai95,Finkelstein97,Wolynes97} but only recently has experimental
data confirmed this prediction
\cite{Galzitskaya03,Li_Pol04,Ivankov04,Kubelka04,Munoz05}.  
It has been shown that $\tau_F$
can be approximately evaluated using $\tau_F \approx \tau_F^0
\exp(N^{\beta})$ where $1/2 \le \beta < 2/3$ with
the prefactor $\tau_F^0$ being on the order
of a $\mu s$.

Much less attention has been paid to finite size effects on the
cooperativity of transition from unfolded states to the
native basin of attraction (NBA). Because 
$N$ is finite large conformational fluctuations are possible which
require careful examination \cite{KlimThirJCC02,Li_Pol04,Li_PRL04,Li_PhysA05}.  For large enough $N$ it is likely that the
folding or melting temperature itself may not be unique
\cite{Holtzer97,Gruebele05,UdgaonkarNSB2001}.  Although substantial
variations in $T_m$ are unlikely it has already been shown that the there
is a range of temperatures over which individual residues in a protein achieve their
native state ordering \cite{Holtzer97}.
  On the other hand, the global cooperativity, as measured by the
dimensionless parameter $\Omega_c$ (see below for definition) has been
shown to scale as \cite{Li_PRL04}
\begin{equation}
\Omega_c \approx N^{\zeta}
\end{equation}
The surprising finding in Eq. (1) requires some discussion. First, the
exponent $\zeta = 1 + \gamma$, where $\gamma$ is the exponent that
describes the divergence of susceptibility at the critical point for a
n-component $\phi ^4$-model with n = 0. It follows that for proteins $\zeta (\approx 2.22)$ is
universal. Second, Eq. (1) is only valid near the folding temperature
$T_F$. At or above $T_F$ the global conformations of the polypeptide
chains as measured by $R_g$ obey the Flory law, i.e. $R_g \approx
a N^{\nu}$ where $\nu \approx 0.6$ \cite{Plaxco04}.
Thus, the unfolded character of the
polypeptide chains are reflected in the thermodynamic cooperativity
of the folding transition at $T_F$.

In this paper we use lattice models with side chains (LMSC),
off-lattice Go models for 23 proteins and
experimental results for a number of proteins to further confirm the
theoretical predictions.  Our results show that $\zeta \approx 2.22$ which
is \textit{distinct from the expected
result} ($\zeta = 2.0$) \textit{for a strong first order transition} \cite{FischerPRB1982}. 
The larger data set of proteins for which folding rates are available
shows that the folding time scales as
\begin{equation}
\tau_F = \tau_0 \exp(cN^{\beta})
\end{equation}
with $c \approx 1.1$, $\beta = 1/2$ and $\tau_0 \approx 0.2 \mu s$.

\section{Models and methods}

{\it Lattice models with side chains (LMSC)}: Each amino acid is represented
using the backbone (B) C$_{\alpha}$ atom that is covalently linked to a unified atom
representing the side chain (SC). Both the C$_{\alpha}$ atoms
and the SCs are
confined to the vertices of a cubic lattice with spacing {\it a}. Thus, a polypeptide chain
consisting of $N$ residues is represented using 2$N$ beads. The energy
of a conformation is 
\begin{eqnarray}
E \; = \;  \epsilon _{bb} \sum_{i=1,j>i+1}^{N} \,
 \delta _{r_{ij}^{bb},a}
+ \epsilon _{bs} \sum_{i=1,j\neq i}^{N} \, \delta _{r_{ij}^{bs},a}
+ \epsilon _{ss} \sum_{i=1,j>i}^{N} \, \delta _{r_{ij}^{ss},a} \; ,
\label{energy_eq}
\end{eqnarray}
where $\epsilon _{bb}, \epsilon _{bs}$ and $\epsilon _{ss}$ are
backbone-backbone(BB-BB), backbone-side chain
(BB-SC) and side chain-side chain (SC-SC) contact energies, respectively.
The distances $r_{ij}^{bb}, r_{ij}^{bs}$ and $r_{ij}^{ss}$  are between
BB, BS and SS beads, respectively. 
The contact energies $\epsilon _{bb}, \epsilon _{bs}$
and $\epsilon _{ss}$ are taken to be -1 (in units of k$_{b}$T) for native 
and 0 for non-native interactions. The neglect of interactions between residues
not present in the native state is the approximation used in the Go model. Because
we are interested in general scaling behavior the use of the Go model is justified.

{\it Off-lattice model}: 
We employ coarse-grained off-lattice models for polypeptide chains
in which each amino acid is represented using only the
C$_{\alpha}$ atoms \cite{HoneyThirBiop92}.
Furthermore, we use a Go model
\cite{Go} in which the interactions between residues forming native
contacts are assumed to be attractive and the non-native
interactions are repulsive. Thus, by definition for the Go model the
PDB structure is the native
 structure with the lowest energy.
The energy of a conformation of the polypeptide chain
 specified by the coordinates $r_i$ of
the C$_{\alpha}$ atoms is \cite{Clementi00}
\begin{eqnarray}
&E& \; = \; \sum_{bonds} K_r (r_{i,i+1} - r_{0i,i+1})^2 + \sum_{angles} K_{\theta}
(\theta_i - \theta_{0i})^2 \nonumber \\
&+& \sum_{dihedral} \{ K_{\phi}^{(1)}
[1 - \cos (\Delta \phi_i)] +
K_{\phi}^{(3)} [1 - \cos 3(\Delta \phi_i)] \} \nonumber\\
&+& \sum_{i>j-3}^{NC}  \epsilon_H \left[ 5 R_{ij}^{12} - 6 R_{ij}^{10}\right] +
\sum_{i>j-3}^{NNC} \epsilon_H \left(\frac{C}{r_{ij}}\right)^{12} .
\label{Hamiltonian}
\end{eqnarray}
Here $\Delta \phi_i=\phi_i - \phi_{0i}$, $R_{ij}={r_{0ij}}/{r_{ij}}$;
$r_{i,i+1}$ is the distance between beads $i$ and $i+1$, $\theta_i$
is the bond angle
 between bonds $(i-1)$ and $i$, 
and $\phi_i$ is the dihedral angle around the $i$th bond and
$r_{ij}$ is the distance between the $i$th and $j$th residues.
Subscripts ``0'', ``NC'' and ``NNC'' refer to the native
conformation, native contacts and non-native contacts,
respectively. Residues $i$ and $j$
are in native contact if $r_{0ij}$ is less than a cutoff
distance $d_c$ taken to be $d_c = 6$ \AA,
where $r_{0ij}$ is the distance between the residues in
the native conformation.

The first harmonic term in Eq. (\ref{Hamiltonian})
accounts for chain
connectivity and the second term represents the bond angle potential.
The potential for the
dihedral angle degrees of freedom is given by the third term in
Eq. (\ref{Hamiltonian}). The interaction energy between residues that are
separated by at least 3 beads is given by 10-12 Lennard-Jones potential.
A soft sphere (last term in Eq. (\ref{Hamiltonian})) repulsive potential
disfavors the formation of non-native contacts.
We choose $K_r =
100 \epsilon _H/\AA^2$, $K_{\theta} = 20 \epsilon _H/rad^2,
 K_{\phi}^{(1)} = \epsilon _H$, and
$K_{\phi}^{(3)} = 0.5 \epsilon _H$, 
where $\epsilon_H$ is the
characteristic hydrogen bond energy and $C = 4$ \AA.

{\it Simulations}: For the LMSC we performed Monte Carlo simulations using
the previously well-tested move set MS3 \cite{Li_JPCB02}. This move set ensures that
ergodicity is obtained efficiently even for $N=50$, it uses
single, double and triple bead moves \cite{Betancourt98}.
Following standard practice the thermodynamic properties are computed
using the multiple histogram method \cite{Ferrenberg89}. The kinetic simulations are carried out
by a quench from high temperature to a temperature at which the NBA
is preferentially populated. The folding times are calculated
from the distribution of first passage times.

For off-lattice models, we assume the dynamics of the polypeptide chain obeys the Langevin
equation. The equations of motion were integrated using the velocity form 
of the Verlet algorithm with the time step $\Delta t = 0.005 \tau_L$,
where $\tau_L = (ma^2/\epsilon_H)^{1/2} \approx 3$ ps.
In order to calculate the thermodynamic quantities we collected
histograms for the energy and native contacts
at five or six different temperatures
(at each temperature 20 - 50 trajectories were generated depending on proteins).
As with the LMSC we used the multiple histogram method \cite{Ferrenberg89} 
to obtain the thermodynamic parameters at all temperatures.

For off-lattice models the probability of being in the native state is computed
using
\begin{equation}
\ f \; = \frac{1}{Q_{\rm T}} \sum_{i<j+1}^N \,\;
\theta (1.2r_{0ij} - r_{ij}) \Delta_{ij} ,
\label{chi_eq}
\end{equation}
where
$\Delta_{ij}$ is equal to 1 if residues $i$ and $j$ form a native
contact and 0 otherwise and, $Q_{\rm T}$ is the total number of
native contacts and $\theta (x)$ is the Heaviside
function.
For the LMSC model we used the structural overlap function \cite{Camacho93PNAS}
\begin{eqnarray}
\chi \; = \;  \frac{1}{2N^{2} - 3N + 1} \left[ \sum_{i<j} \,
 \delta (r_{ij}^{ss} - r_{ij}^{ss,N})
+ \sum_{i<j+1} \, \delta (r_{ij}^{bb} - r_{ij}^{bb,N})
+ \sum_{i \neq j} \, \delta (r_{ij}^{bs} - r_{ij}^{bs,N}) \; \right].
\label{energy_eq}
\end{eqnarray}
The overlap function $\chi$, which is one if the
conformation of the polypeptide chain coincides with the native
structure and is small for unfolded conformations, is an
order parameter for the folding-unfolding transition. The
probability of being in the native state $f_N$
is $f_N = <f> = 1 - <\chi>$, where $<...>$ denotes a thermal average.

{\it Cooperativity}.
The extent of cooperativity of the transition to the NBA from the ensemble of
unfolded states is measured using the dimensionless parameter
\begin{equation}
\Omega _c \; = \; \frac{T_F^2}{\Delta T}
\left|\frac{df_N}{dT}\right|_{T=T_F} ,
\label{coop_eq}
\end{equation}
where $\Delta T$ is the full width at half-maximum of $df_N/dT$ and
the folding temperature $T_F$ is identified with the maximum of 
$df_N/dT$. Two points about $\Omega_c$ are noteworthy. (1) For
proteins that melt by a two-state transition it is trivial  to show that
$\Delta H_{vH} = 4k_B\Delta T\Omega _c$, where $\Delta H_{vH}$ is the
van't Hoff enthalpy at $T_F$. For an infinitely sharp two-state transition
there is a latent heat release at $T_F$, at which $C_p$ can be approximated
by a delta-function. In this case $\Omega_c \rightarrow \infty$ which implies
that $\Delta H_{vH}$ and the calorimetric enthalpy $\Delta H_{cal}$
(obtained by integrating the temperature dependence of the specific heat
$C_p$ ) would coincide. It is logical to infer
that as $\Omega_c$ increases the ratio $\kappa = \Delta H_{vH}/\Delta H_{cal}$
should approach unity. 
(2)  Even for moderate sized proteins that undergo a two-state transition
$\kappa \approx 1$ \cite{Privalov79}.
It is known that the extent of cooperativity depends on external
conditions as has been demonstrated for thermal denaturation of CI2 at
several values of pH \cite{Jackson91}. The values of $\kappa$ for all
pH values are $\approx 1$.
However, the variation in cooperativity of CI2 as pH varies are
reflected in the changes in  $\Omega _c$ \cite{KlimThirum98FD}.
Therefore, we believe that $\Omega _c$, that varies in the
range $0 < \Omega _c < \infty$, is a better descriptor of the extent of
cooperativity than $\kappa$. The latter merely tests the applicability
of the two-state approximation.
 
\section{Results}

\subsection{Dependence of $\Omega_c$ on $N$}

For the 23 Go proteins listed in Table I, we calculated $\Omega_c$ from
the temperature dependence of $f_N$. In Fig. \ref{hairpin_CspB_fig}
we compare the temperature dependence of $f_N(T)$ and $df_N(T)/dT$ for
$\beta$-hairpin ($N=16$) and {\it Bacillus subtilis} (CpsB, $N=67$). 
It is clear that the transition width and the amplitudes of $df_N/dT$ 
obtained using Go models, compare only qualitatively well with experiments.
As pointed out by Kaya and Chan \cite{Kaya00,Kaya03,Chan_ME04,ChanPRL2000},
the simple Go-like models consistently
underestimate the extent of cooperativity. Nevertheless, both the models and
experiments show that $\Omega_c$ increases dramatically as $N$ increases
(Fig. \ref{hairpin_CspB_fig}).

The variation of $\Omega_c$ with $N$ for the 23 proteins obtained from
the simulations of Go models is given in Fig. \ref{Scal_Omega_fig}.
From the ln$\Omega_c$-ln$N$ plot we obtain $\zeta = 2.40 \pm 0.20$
and $\zeta = 2.35 \pm 0.07$ for off-lattice models and LMSC, respectively. These
values of $\zeta$ deviate  from the theoretical prediction
$\zeta \approx 2.22$.
We suspect that this is due to large fluctuations in the native state of
polypeptide chains that are represented using minimal models. 
Nevertheless, the results for the minimal models rule out
the value of $\zeta = 2$ that is predicted for systems that undergo first
order transition. The near coincidence of $\zeta$ for both models show that
the details of interactions are not relevant.

For the thirty four proteins (Table II) for which we could find thermal
denaturation data, we calculated $\Omega_c$ using the $\Delta H$, 
and $T_F$ (referred to as the melting temperature $T_m$ in the experimental
literature). From the plot of ln$\Omega_c$ versus ln$N$ we find that
$\zeta = 2.17 \pm 0.09$. The experimental value of $\zeta$, which also
deviates from $\zeta = 2$, is in much better agreement with the theoretical
prediction. The analysis of experimental data requires care because the
compiled results were obtained from a number of different laboratories around 
the world. Each laboratory uses different methods to analyze the raw 
experimental data which invariably lead to varying methods to
estimate errors in
$\Delta H$ and $T_m$. To estimate the error bar for $\zeta$ it is important
to consider the errors in the computation of $\Omega_c$. 
Using the reported experimental errors in $T_m$ and
$\Delta H$ we calculated the variance $\delta^2\Omega_c$ using the standard 
expression for the error propagation \cite{Li_PRL04,Webpage}.
%\begin{equation}
%\delta^2\Omega_c \; = \; \left(\frac{\partial \Omega_c}{\partial T_m}\right)^2
%\delta^2T_m + \left(\frac{\partial \Omega_c}{\partial \Delta H}\right)^2
%\delta^2\Delta H + 2 \left|\frac{\partial \Omega_c}{\partial T_m}
%\frac{\partial \Omega_c}{\partial \Delta H}\right|\delta T_m\delta\Delta H,
%\label{error_eq}
%\end{equation}
The upper bound in the error
in $\Omega_c$ 
for the thirty four proteins is given in Table II. To provide an accurate 
evaluation of the errors in the exponent $\zeta$ we used a weighted linear
fit, in which each value of ln$\Omega_c$ contributes to the fit with the
weight proportional to its standard deviation \cite{Li_PRL04, Webpage}.

\subsection{Dependence of folding free energy barrier on $N$}

The simultaneous presence of stabilizing (between hydrophobic residues) and
destabilizing interactions involving polar and charged residues in 
polypeptide chain renders the native state only marginally stable
\cite{Poland}.
The hydrophobic residues enable the formation of compact structures while
polar and charged residues, for whom water is a good solvent, are better
accommodated by extended conformations. Thus, in the folded state the
average energy gain per residue (compared to expanded states) is 
$-\epsilon _H (\approx (1 - 2)$ kcal/mol) whereas due to chain connectivity
and surface area burial the loss in free energy of exposed residues is
$\epsilon _P \approx \epsilon _H$. Because there is a large number of
solvent-mediated interactions that stabilize the native state,
 even when $N$ is small, it follows from the
central limit theorem that the barrier height $\beta \Delta G^{\ddagger}$,
whose lower bound is the stabilizing free energy should scale as
$\Delta G^{\ddagger} \sim k_BT\sqrt{N}$ \cite{Thirumalai95}.
A different physical picture has been used to argue that 
$\Delta G^{\ddagger} \sim k_BTN^{2/3}$ \cite{Finkelstein97,Wolynes97}.
Both the scenarios show that the barrier to folding rates 
scales sublinearly with $N$. 

The dependence of ln$k_F$ ($k_F = \tau_F^{-1}$) on $N$ using experimental
data for 69 proteins \cite{Munoz05}
and the simulation results for the 23 proteins is
consistent with the predicted behavior that
$\Delta G^{\ddagger} = ck_BT\sqrt{N}$ with $c \approx 1$. The correlation
between the experimental results and the theoretical fit is 0.74
which is similar to the previous analysis using a set of 57
proteins \cite{Li_Pol04}. It should be noted that the data can also be fit using
$\Delta G^{\ddagger} \sim k_BTN^{2/3}$. 
The prefactor $\tau_F^0$ using the $N^{2/3}$ fit is over
an order of magnitude larger than
for the $N^{1/2}$ behavior. In the absence of accurate
measurements for a larger data set of proteins it is difficult to
distinguish between the two power laws for $\Delta G^{\ddagger}$.

Previous studies \cite{KlimThirJCP98,Wolynes92} have shown that there is a correlation between folding
rates and $Z$-score which can be defined as
\begin{equation}
Z_G \; = \; \frac{G_N - <G_U>}{\sigma} ,
\label{Zscore_eq}
\end{equation}
where $G_N$ is the free energy of the native state, $<G_U>$ is the average free
energy of the unfolded states and $\sigma$ is the dispersion in the free
energy of the unfolded states. From the fluctuation formula it follows that
$\sigma = \sqrt{k_BT^2C_p}$ so that
\begin{equation}
Z_G \; = \; \frac{\Delta G}{\sqrt{k_BT^2C_p}} .
\label{Zscore1_eq}
\end{equation}
Since $\Delta G$ and $C_p$ are extensive it follows that $Z_G \sim N^{1/2}$.
This observation establishes an intrinsic connection between the
thermodynamics and kinetics of protein folding that involves formation and
rearrangement of non-covalent interactions. In an interesting
 recent note \cite{Munoz05}
it has been argued that the finding
 $\Delta G^{\ddagger} \sim k_BT\sqrt{N}$ can be
interpreted in terms of $n_{\sigma}$ in which $\Delta G$ in
 Eq. (\ref{Zscore1_eq}) is replaced by $\Delta H$. In either case, there
appears to be a thermodynamic rationale for the sublinear scaling
of the folding free energy barrier.

\section{Conclusions}

We have reexamined the dependence of the extent of cooperativity as a function
of $N$ using lattice models with side chains, off-lattice models and experimental data on thermal denaturation.
The finding that $\Omega _c \sim N^{\zeta}$ at $T \approx T_F$ with $\zeta > 2$
provides additional support for the earlier theoretical predictions \cite{Li_PRL04}. More
importantly, the present work also shows that the theoretical value for
$\zeta$ is independent of the precise model used which implies that $\zeta$
is universal. It is surprising to find such general characteristics for
proteins for which specificity is often an important property. We should note
that accurate value of $\zeta$ and $\Omega _c$ can only be obtained using
more refined models that perhaps include desolvation 
penalty \cite{Kaya03,Cheung02}

In accord with a number of theoretical predictions
\cite{Thirumalai95,Finkelstein97,Wolynes97,Shakhnovich96,Li_JPCB02,TakadaJMB2001}
we found that the folding free energy barrier scales only sublinearly
with $N$. The relatively small barrier is in accord with the marginal stability
of proteins. Since the barriers to global unfolding is relatively small it
follows that there must be large conformational fluctuations even when the
protein is in the NBA. Indeed, recent experiments show that such dynamical
fluctuations that are localized in various regions of a monomeric protein might
play an important functional role. These observations suggest that small barriers in proteins and RNA \cite{Hyeon05}
might be an evolved characteristics of all natural sequences.

This work was
supported in part by a KBN grant  No 1P03B01827,
the National Science Foundation grant
(CHE 05-14056) and National Science Council in Taiwan under grant numbers
No. NSC 93-2112-M-001-027 (to CKH).

%\newpage
%
%\section*{\bf Table Caption}
%
%{\bf Table 1}: List of 23 proteins used in the simulations.
%(a) The native state for use in the Go model is obtained from the structures deposited in the Protein Data Bank. (b) $\Omega _c$ is calculated 
%using equation (\ref{coop_eq}) with $f_N = <\chi (T)>$. 
%(c) We calculated $\delta \Omega _c$ by the histogram method.
%
%{\bf Table 2}: List of 34 proteins for which $\Omega _c$ is calculated
%using experimental data. The calculated $\Omega _c$ values from experiments 
%are significantly larger than those obtained using the Go models (see Table 1).

\newpage

\begin{table}[h]
\begin{tabular}{|c|c|c|c|c|}
\hline
Protein&$N$&PDB code$^{\rm a}$&$\Omega_c^{\rm b}$&$\delta \Omega_c^{\rm c}$ \\
\hline
$\beta$-hairpin&$16$&1PGB&2.29&0.02\\
\hline
$\alpha$-helix&$21$&no code&0.803&0.002\\
\hline
WW domain&$34$&1PIN&3.79&0.02\\
\hline
Villin headpiece&$36$&1VII&3.51&0.01\\
\hline
YAP65&$40$&1K5R&3.63&0.05\\
\hline
E3BD&$45$& &7.21&0.05 \\
\hline
hbSBD&$52$&1ZWV&51.4&0.2\\
\hline
Protein G&$56$&1PGB&16.98&0.89\\
\hline
SH3 domain ($\alpha$-spectrum)&$57$&1SHG&74.03&1.35\\
\hline
SH3 domain (fyn)&$59$&1SHF&103.95&5.06\\
\hline
IgG-binding domain of streptococcal protein L&$63$&1HZ6&21.18&0.39\\
\hline
Chymotrypsin Inhibitor 2 (CI-2)&$65$&2CI2&33.23&1.66\\
\hline
CspB (Bacillus subtilis)&$67$&1CSP&66.87&2.18\\
\hline
CspA&$69$&1MJC&117.23&13.33\\
\hline
Ubiquitin&$76$&1UBQ&117.8&11.1\\
\hline
Activation domain procarboxypeptidase A2&$80$&1AYE&73.7&3.1\\
\hline
His-containing phosphocarrier protein&$85$&1POH&74.52&4.2\\
\hline
hbLBD&$87$&1K8M&15.8&0.2\\
\hline
Tenascin (short form)&$89$&1TEN&39.11&1.14\\
\hline
Twitchin Ig repeat 27&$89$&1TIT&44.85&0.66\\
\hline
S6&$97$&1RIS&48.69&1.31\\
\hline
FKBP12&$107$&1FKB&95.52&3.85\\
\hline
Ribonuclease A&$124$&1A5P&69.05&2.84\\
\hline
\end{tabular}
\caption{List of 23 proteins used in the simulations.
(a) The native state for use in the Go model is obtained from the structures deposited in the Protein Data Bank. (b) $\Omega _c$ is calculated
using equation (\ref{coop_eq}) with $f_N = <\chi (T)>$.
(c) 2 $\delta \Omega _c = |\Omega _c - \Omega _{c_1}| + |\Omega _c - \Omega _{c_2}|$, where $\Omega _{c_1}$ and $\Omega _{c_2}$ are
values of the cooperativity measure obtained by retaining only one-half the conformations used to compute $\Omega _c$.}
\end{table}

\newpage

\begin{table}[h]
\begin{tabular}{|c|c|c|c|c|c|c|c|c|}
\hline
Protein&$N$&$\Omega_c^a$&$\delta \Omega_c^b$& &Protein&$N$&$\Omega_c^a$&$\delta \Omega_c^b$ \\
%\hline
\cline{1-4} \cline{6-9}
BH8 $\beta$-hairpin \cite{Dyer}&12&12.9&0.5& &SS07d \cite{Knapp1}&64&555.2&56.2\\
%\hline
\cline{1-4} \cline{6-9}
HP1 $\beta$-hairpin \cite{Gai}&15&8.9&0.1& &CI2 \cite{Jackson91}&65&691.2&17.0  \\
%\hline
\cline{1-4} \cline{6-9}
MrH3a $\beta$-hairpin \cite{Dyer}&16&54.1&6.2& &CspTm \cite{Jaenicke} &66&558.2&56.3 \\
%\hline
\cline{1-4} \cline{6-9}
$\beta$-hairpin \cite{Honda}&16&33.8&7.4& &Btk SH3 \cite{Knapp2} &67&316.4&25.9\\
%\hline
\cline{1-4} \cline{6-9}
Trp-cage protein \cite{Hagen}&20&24.8&5.1& &binary pattern protein \cite{Hechts} &74&273.9&30.5 \\
%\hline
\cline{1-4} \cline{6-9}
$\alpha$-helix \cite{Williams96}&21&23.5&7.9& &ADA2h \cite{Villegas95}  &80&332.0&35.2\\
%\hline
\cline{1-4} \cline{6-9}
villin headpeace \cite{Kubelka03}&35&112.2&9.6& &hbLBD \cite{Huang04} &87&903.1&11.1 \\
%\hline
\cline{1-4} \cline{6-9}
FBP28 WW domain$^c$ \cite{Ferguson01}&37&107.1&8.9& &tenascin Fn3 domain \cite{Clarke97} &91&842.4&56.6\\
%\hline
\cline{1-4} \cline{6-9}
FBP28 W30A WW domain$^c$ \cite{Ferguson01} &37&90.4&8.8& &Sa RNase \cite{SaRNase} &96&1651.1&166.6 \\
%\hline
\cline{1-4} \cline{6-9}
WW prototype$^c$ \cite{Ferguson01}&38&93.8&8.4& &Sa3 RNase \cite{SaRNase}&97&852.7&86.0\\
%\hline
\cline{1-4} \cline{6-9}
YAP WW$^c$ \cite{Ferguson01}&40&96.9&18.5& &HPr \cite{VanNuland98}&98&975.6&61.9 \\
%\hline
\cline{1-4} \cline{6-9}
BBL \cite{Ferguson_p}&47&128.2&18.0& &Sa2 RNase \cite{SaRNase} &99&1535.0&156.9 \\
%\hline
\cline{1-4} \cline{6-9}
PSBD domain \cite{Ferguson_p}&47&282.8&24.0& &barnase \cite{barnase}&110&2860.1&286.0 \\
%\hline
\cline{1-4} \cline{6-9}
PSBD domain \cite{Ferguson_p}&50&176.2&13.0& &RNase A \cite{Arnold97}&125&3038.5&42.6 \\
%\hline
\cline{1-4} \cline{6-9}
hbSBD \cite{Kouza05} &52&71.8&6.3& &RNase B \cite{Arnold97}&125&3038.4&87.5\\
%\hline
\cline{1-4} \cline{6-9}
B1 domain of protein G \cite{Alexander92} &56&525.7&12.5& &lysozyme \cite{Hirai} &129&1014.1&187.3 \\
%\hline
\cline{1-4} \cline{6-9}
B2 domain of protein G \cite{Alexander92} &56&468.4&20.0& &interleukin-1$\beta$ \cite{Privalov} &153&1189.6&128.6\\\hline
\end{tabular}
\caption{List of 34 proteins for which $\Omega _c$ is calculated
using experimental data. The calculated $\Omega _c$ values from experiments
are significantly larger than those obtained using the Go models (see Table 1).
a) $\Omega _c$ is computed at $T = T_F = T_m$ using the experimental values
of $\Delta H$ and $T_m$.
b) The error in $\delta \Omega_c$ is computed using the proceedure given in \cite{Li_PRL04,Shakhnovich96}.
c) Data are averaged over two salt conditions at pH 7.0.}
\end{table}

\newpage
\section*{\bf Figure Caption}

\begin{figure}[ht]
\includegraphics[width=6.00in]{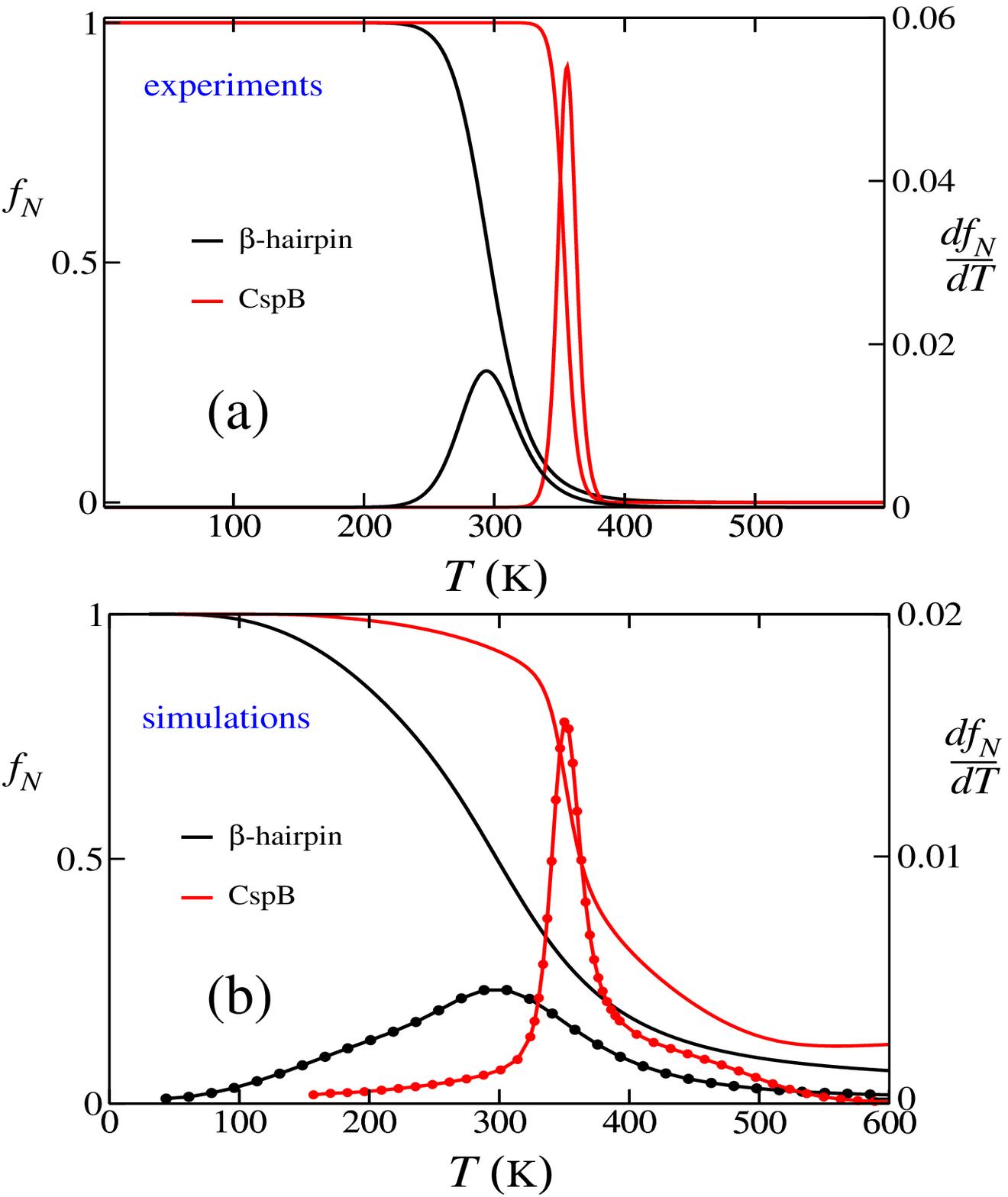}
\caption{}
\label{hairpin_CspB_fig}
\end{figure}

{{\bf Figure} \ref{hairpin_CspB_fig} :}
The temperature dependence of $f_N$ and $df_N/dT$ for $\beta$-hairpin
($N=16$) and CpsB ($N=67$). The scale for $df_N/dT$ is given on the right.
(a): the experimental curves were
obtained using
$\Delta H = 11.6$ kcal/mol,
$T_m=297$ K and $\Delta H = 54.4$ kcal/mol and $T_m= 354.5$ K for
$\beta$-hairpin and CpsB, respectively.
(b): the simulation results were calculated from $f_N = <\chi (T)>$.
The Go model gives only a qualitatively reliable estimates of $f_N(T)$.

{{\bf Figure} \ref{Scal_Omega_fig}:}
Plot of ln$\Omega_c$ as a function of ln$N$.
The red line is a  fit to the simulation data for the 23
off-lattice Go proteins from  which we estimate
$\zeta =2.40 \pm 0.20$. The black line is a fit to the lattice models
with side chains ($N = 18, 24, 32, 40$ and 50) with
$\zeta = 2.35 \pm 0.07$.
The blue line is a fit to the experimental values of
$\Omega_c$ for 34 proteins (Table 2)
with $\zeta = 2.17 \pm 0.09$. The larger deviation in $\zeta$ for the minimal
models is due to lack of all the interactions that stabilize the native state .

\begin{figure}[ht]
\includegraphics[width=6.0in]{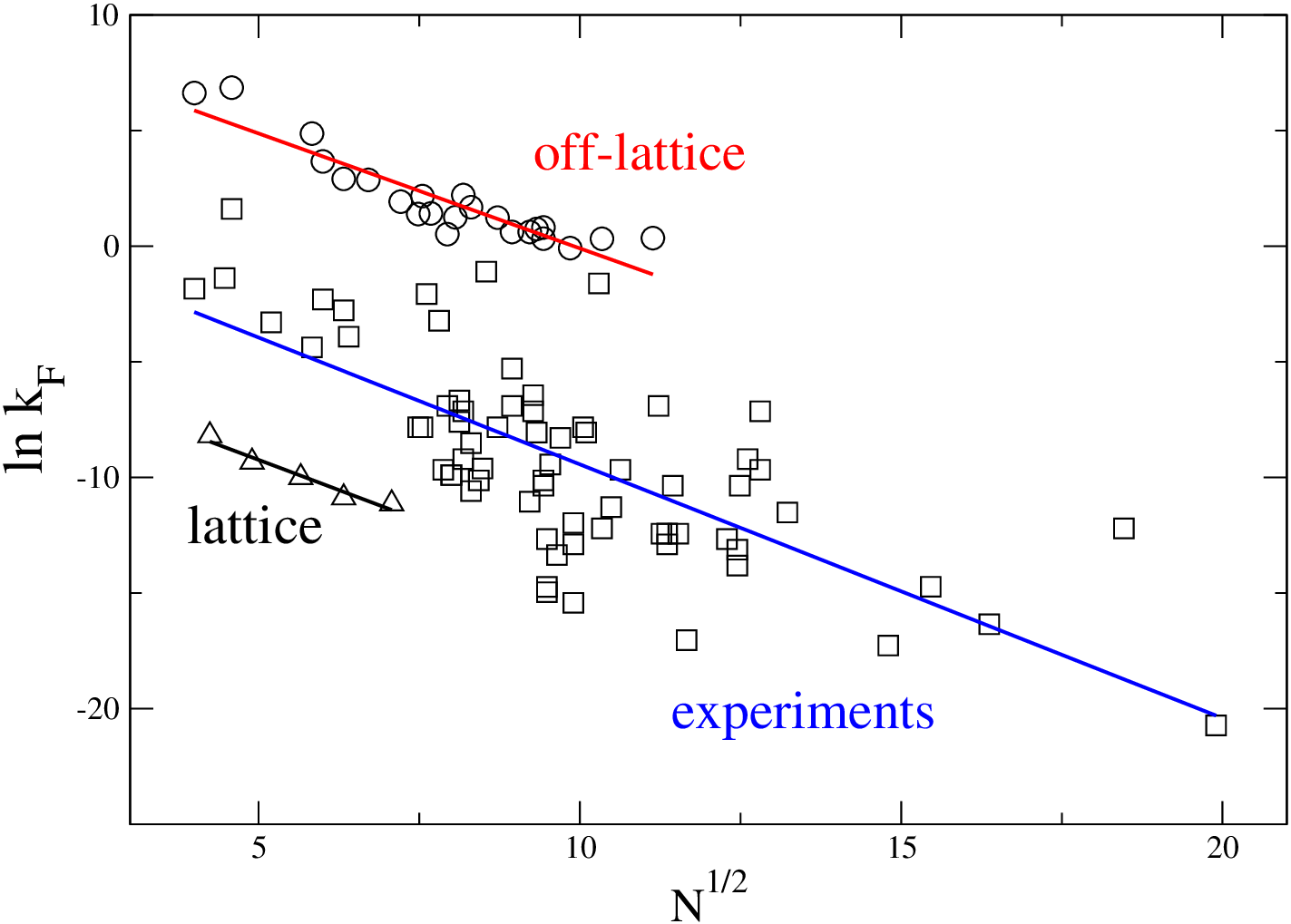}
\caption{}
\label{real_pro_fig}
\end{figure}

{{\bf Figure} \ref{real_pro_fig}:}
Folding rate of 69 real proteins (squares) is plotted as
a function of $N^{1/2}$ (the straight line represent the fit
$y = 1.54 -1.10x$ with the correlation coefficient $R=0.74$).
The open circles represent the data obtained for 23
off-lattice Go proteins (see Table 1)
(the linear fit $y = 9.84 - x$ and $R=0.92$).
The triangles denote the data obtained for lattice models with side
chains ($N = 18, 24, 32, 40$ and 50, the linear fit
$y = -4.01 - 1.1x$ and $R=0.98$). For real proteins
and off-lattice Go proteins $k_F$ is measured in $\mu s^{-1}$, whereas
for the lattice models it is measured in MCS$^{-1}$
where MCS is Monte Carlo steps.

\newpage

% FIGURE 1
%\begin{figure}
%\epsfxsize=3.2in
%\centerline{\epsffile{fig1_royal.eps}}

\newpage

% FIGURE 2
%\epsfxsize=5.5in
%\centerline{\epsffile{fig2_royal.eps}}
\begin{figure}[ht]
\includegraphics[width=6.00in]{fig2_new.eps}
\caption{}
\label{Scal_Omega_fig}
\end{figure}

% FIGURE 3
%\begin{figure}
%\epsfxsize=3.2in
%\centerline{\epsffile{fig3_royal.eps}}

\end{document}